\documentclass[10pt,aps,prd,reprint,nofootinbib,floats,floatfix,amsfonts,amssymb,amsmath,preprintnumbers,notitlepage,superscriptaddress]{revtex4-1}
\usepackage{graphicx}
\usepackage[utf8]{inputenc}
\usepackage[british]{babel}
\usepackage{isomath}
\usepackage[protrusion=true,tracking=true,kerning=true,spacing=true,final,babel=true]{microtype}
\usepackage{physics}
\usepackage{xcolor}

\usepackage[final]{hyperref}
\usepackage{nth}

\begin{document}

\begin{flushleft}
KCL-PH-TH/2019-67
\end{flushleft}

\title{Constraints on Quasi-dilaton Massive Gravity}
\author{Katarina~Martinovic}
\email{katarina.martinovic@kcl.ac.uk}
\author{Mairi~Sakellariadou}
\email{mairi.sakellariadou@kcl.ac.uk}
\affiliation{Theoretical Particle Physics and Cosmology Group, \, Physics \, Department, \\ King's College London, \, University \, of London, \, Strand, \, London \, WC2R \, 2LS, \, UK}
%\textit{\scriptsize{\parbox{11cm}{\centering Theoretical Particle Physics and Cosmology Group, Physics Department, King's College London, University of London, Strand, London \, WC2R 2LS, UK}}}
\date{\today}    

\begin{abstract}
   The last decade has seen great advancements in the field of modified gravity, motivated by the dark energy problem, or by the search for a fundamental quantum gravity theory. With a phenomenologically-driven analysis, we consider dRGT theory and its extension, quasi-dilaton massive gravity (QDMG). When looking for ways to constrain the theory, a promising direction appeared to be astrophysical tests. The scalar gravitational degree of freedom and quasi-dilaton degree of freedom alter the evolution of Bardeen potentials, which in turn affects the galaxy rotation curves. We find an upper bound on graviton mass in QDMG to be $m \leq 10^{-31} {\rm eV}$. This result agrees with bounds from LIGO and numerous Solar System tests. However, the extremely small mass of the graviton remains a detection out of reach, with LISA's sensitivity exploring the parameter space up to $m \leq 10^{-25} {\rm eV}$.
\end{abstract}

\maketitle

\section{Introduction}

General Relativity (GR) has exhibited immense success in Solar System tests \cite{Reynaud:2008yd}, as well as weak field \cite{Collett:2018gpf} and strong field regimes (e.g. merger events \cite{TheLIGOScientific:2016src} and pulsars \cite{Kramer:2006nb}), in recent years. While the success of General Relativity strongly suggests that this theory is indeed a good description of gravity, exploring modifications to it is an important test of the theory itself. Moreover, a modification of GR may provide a natural explanation of the current accelerated expansion of the Universe without the need of a dark energy component \cite{PhysRevD.71.063513}. \\

The widely celebrated gravitational wave event, GW170817, which detected two neutron stars merging, not only in the gravitational spectrum \cite{TheLIGOScientific:2017qsa}, but in the electromagnetic spectrum too, has had resounding consequences on the field of modified gravity. Comparing the time delay between the gravitational and the electromagnetic signal placed a stringent constraint on the difference of the speeds of the two to be less than 1 part in $10^{-15}$ \cite{Monitor:2017mdv}. This constraint ruled out theories of gravity that predicted a significant deviation of the propagation speed of gravitational waves from the speed of light \cite{Baker:2017hug}, \cite{Ezquiaga:2017ekz}, \cite{PhysRevLett.119.251303}. However, several other candidates of modified gravity remain so far valid. The challenge and important task is to identify the most promising surviving theories and find ways to test them given the currently available astrophysical data.\\ 

Massive gravity has gained increasing interest over the past years due to the works done by de Rham, Gabadadze and Tolley to formulate a ghost-free theory of massive gravity  (dRGT) \cite{deRham:2010kj}. In this paper we look for ways to test an extension of dRGT theory: quasi-dilaton massive gravity (QDMG). QDMG is a dRGT theory with an additional quasi-dilaton scalar field \cite{DAmico:2012hia}.\\ 

In Section 2 we motivate this specific choice of a modified gravity theory and set-up the theoretical framework in which we work. In Section 3 we consider two different possible approaches to constraining the theory. The first approach is motivated by the ever-growing data in the gravitational waves sector and looks at depletion of a gravitational wave signal in a massive gravity theory. The second approach considers astrophysical tests of QDMG, in particular changes to the form of the Bardeen potentials. This in turn leads to modifications to the rotation curves of the theory. Finally, we discuss our results and comment on possible future work in Section 4.

\section{Theoretical Motivation}

In deciding which path to take in the forest of modified gravity theories, we follow a general principle of being interested in top down approaches over bottom up ones. Namely, modifications to the Einstein-Hilbert action should be a consequence of some underlying fundamental theory. Additionally, we wanted new degrees of freedom of the theory to be intrinsically gravitational, rather than adding new fields on top of General Relativity. Keeping this in mind, as a first step, we pursued work in dRGT theory. As will be discussed in the later sections of the paper, we ended up needing to add a new degree of freedom on top of dRGT, in order to obtain a stable cosmology.\\

The naive thing to do when modifying GR is to add a covariant mass term to the Einstein-Hilbert action. This produces inconsistent results because the massless limit of this theory does not recover gravity, an inconsistency known in the literature as vDVZ (discussed by van Dam, Veltman, and Zakharov \cite{vanDam:1970vg,Zakharov:1970cc}) discontinuity. It arises due to the presence of extra gravitational degrees of freedom, some of which couple to matter \cite{Babichev:2013usa, Hinterbichler:2011tt}. To resolve the problem, one must also introduce additional non-linear terms that screen these new degrees of freedom, a mechanism referred to as Vainshtein screening \cite{PhysRevD.83.103516}.
The theoretical formulation of dRGT gravity was a turning point for all development in the field of massive spin-2 gravity, as it is the first fully complete ghost-free theory of massive gravity. In addition, with its origin in extra dimensional braneworlds, dRGT allows one to entertain even the possibility of a UV completion of the theory \cite{Cheung:2016yqr}.\\ 

As promising as the theoretical advancements in dRGT are, its phenomenology, however, makes it difficult to practically constrain the theory. It passes all hitherto proposed tests and agrees with General Relativity, making the two indistinguishable. Out of the additional degrees of freedom, the vector modes do not interact with matter, and the scalar mode is Vainshtein-screened close to a source which significantly represses both its production and detection \cite{deRham:2014zqa}. In addition, a great concern in searching for tests of the theory is the fact that dRGT gravity does not support stable Friedman-Lema\^{i}tre-Robertson-Walker (FLRW) solutions. Hence, to promote dRGT to a cosmological theory, one has to add extra degrees of freedom, be it in the form of tensorial modes (e.g. bigravity \cite{PhysRevD.99.104032}) or scalar modes (e.g. f(R) massive gravity \cite{Cai:2014upa}). \\

In what follows, we focus on quasi-dilaton massive gravity, which is a scalar-extended dRGT theory with a quasi-dilaton field leading to stable FLRW solutions. We note that the existence of a quasi-dilaton type of field is well-founded within string theory and it arises from compactification of the extra dimensions \cite{Damour:1994zq}.\\

The action of quasi-dilaton massive gravity (QDMG) is

\begin{eqnarray}
    \mathcal{S}_{\text{QDMG}}& =& \int \text{d}^4 x \sqrt{-g} \Big[-\frac{M^2_{\rm{Pl}}}{2} R(g) + \frac{\omega}{2} (\partial \sigma)^2\nonumber\\
    &&+ m^2M^2_{\rm{Pl}} \sum^4_{n=0} \alpha_n \mathcal{L}_n( \mathcal{K}(g,f,\sigma))
    \nonumber\\
    &&+ \;\mathcal{L}_m (g, \Phi_i)\Big],
\end{eqnarray}
where $\sigma$ is the quasi-dilaton, $g$ is the dynamical metric, $f$ is the Stückelbergised fiducial metric, and we define $ \mathcal{K}$ in terms of the Stückelberg fields, $\phi^a$, as following:
\begin{equation}
   \mathcal{K}_{\nu}^{\mu} = \delta_{\nu}^{\mu} - e^{\sigma/M_{\text{Pl}}}\sqrt{g^{\mu\alpha}\partial_{\alpha}\phi^a\partial_{\nu}\phi^b \eta_{ab}}.
\end{equation}
It is the interaction between the dynamical metric, fiducial metric and the quasi-dilaton field that gives rise to the graviton mass, $m$. The functions $\mathcal{L}_n$ are defined in detail in the Appendix. Note that the fiducial metric enters the lagrangian only through the interaction term with the dynamical metric. In other words, it does not directly couple to matter fields, $\Phi_i$, and it does not explicitly affect the geodesics.\\

In the same way as in dRGT theory, to make calculations simpler, one can take the decoupling limit of the above action \cite{Ondo:2013wka}. The lagrangian in the decoupling limit is derived by keeping the so-called decoupling scale, $\Lambda^3 = m^2 M_{\rm pl}$, constant, and letting $M_{\rm pl} \rightarrow \infty$ and $m \rightarrow 0$. One can think of it as an expansion in $1/M_{\rm pl}$, with the relative contribution of the terms in the expansion determined by the above-mentioned scaling relations. Note that in the case $M_{Pl} \rightarrow \infty$, the dynamical metric reduces to Minkowski ($g_{\mu\nu}=\eta_{\mu\nu} + M_\text{Pl}^{-1} h_{\mu\nu} \rightarrow \eta_{\mu\nu}$), and the Einstein-Hilbert term reduces to its linear form.  From here it is clear that the decoupling limit is not appropriate if we are considering non-linear gravity regimes. If, however, we are looking at weak field gravity systems, this is an acceptable simplification \cite{Dar:2018dra}. Note that in our analysis we ignore the vector gravitational degrees of freedom since these do not couple to matter. Interestingly, taking the decoupling limit of QDMG leads to a bi-Galileon theory of gravity, i.e. the scalar gravitational field and the quasi-dilaton field both acquire a galilean symmetry \cite{DAmico:2012hia}.\\

The Lagrangian of the theory of interest (i.e in the decoupling limit (DL)) then reads 
\begin{eqnarray}
\label{eq:lag_qdmg}
    \mathcal{L}^{\text{DL}}_{\text{QDMG}} &=& - \frac{1}{4}\Big( h^{\mu\nu}\hat{\mathcal{E}}^{\alpha\beta}_{\mu\nu}h_{\alpha\beta}+\sum^5_{n=2} \frac{c_n}{\Lambda^{3(n-2)}_3} \mathcal{L}^{(n)}_{\text{Gal}}[\pi]\nonumber\\
    &&- \frac{2(\alpha_3+4\alpha_4)}{\Lambda^6_3}h^{\mu\nu}X^{(3)}_{\mu\nu}[\Pi]\Big)  
    - \frac{\omega}{2} (\partial \sigma)^2\nonumber\\
    && + \frac{1}{2} \sigma \sum^4_{n=1}\frac{(4-n)\alpha_n-(n+1)\alpha_{n+1}}{\Lambda^{3(n-1)}_3} \mathcal{L}_n[\Pi]\nonumber\\
    &&  + \frac{1}{2 M_{\text{Pl}}} h_{\mu\nu}T^{\mu\nu}+ \frac{1}{2 M_{\text{Pl}}}  \pi T\nonumber\\
    &&- \frac{2+3\alpha_3}{4M_{\text{Pl}} \Lambda^3_3} \partial_{\mu}\pi \partial_{\nu}\pi T^{\mu\nu}, 
\end{eqnarray}
where $\Lambda_3$ is the decoupling scale (same as for dRGT), $\pi$ is the gravitational scalar degree of freedom and $\sigma$ is the quasi-dilaton field as before. The explicit form of  $\mathcal{L}^{(n)}_{\text{Gal}}[\pi]$, $X^{(3)}_{\mu\nu}[\Pi]$ and $\mathcal{L}_n[\Pi]$ can be found in the Appendix. The decoupling limit is valid for typical scales bigger than $1/m$,  where $m$ is the graviton mass, and in QDMG we expect this to be of the order of the Hubble scale, $m \approx H_0$ \cite{Gannouji:2013rwa}. Note that all of the fields have been canonically normalised. \\

In the next section we explore two approaches that may constrain quasi-dilaton massive gravity. We firstly estimate the decay probability of helicity-2 to helicity-0 modes, and find that the decay width is too small to leave a trace in the gravitational wave signal.  
%The main result of this paper is a constraint placed on graviton mass in QDMG. 
We then investigate the effect that QDMG could have on astrophysical scales, e.g. galaxies and clusters of galaxies. In particular, we investigate whether we can constrain QDMG and its parameters from rotation curves and gravitational lensing. We note that a similar analysis has been done for beyond-Horndeski theories in \cite{Jain:2015edg, Koyama:2015oma} and more recently in \cite{Salzano:2017qac}. Throughout this paper we use natural units, $c = \hbar = 1$, and the metric signature $(-,+,+,+)$.

\section{Constraints from Data}

\subsection{Depletion of the gravitational wave signal}

One of the unsurprising consequences of working with a massive, instead of a massless, theory of gravity is that the dispersion relation for gravitational waves gets modified. The correction to the dispersion relation in QDMG looks similar to that in dRGT theory where the tensor mode acquires a mass contribution that is of the order of the mass of the graviton \cite{Gumrukcuoglu:2011zh}. Current graviton mass bounds put an upper constraint on the mass, $m < 10 ^ {-22}$ eV \cite{LIGOScientific:2019fpa, deRham:2016nuf}. \\

Here we investigate the possibility of depletion of the gravitational wave signal due to the decay of tensor modes to scalar ones. Our results agree with the analysis in \cite{Creminelli:2018xsv}, where it was showed that higher derivative corrections in the Horndeski effective field theory of dark energy are too small to modify the gravitational wave signal. Working in the decoupling limit of QDMG, outside the Vainsthein screening region, the Lagrangian reduces to that of linearised massive gravity, since the waves mostly travel through vacuum \cite{deRham:2012fw}. Validity of the decoupling limit for gravitational wave signals detected by LIGO/Virgo is under debate, since the decoupling scale is close to the energies observed by the mentioned detectors \cite{deRham:2018red}. LISA, however, will probe scales that are well below the decoupling scale, and therefore the use of the decoupling limit is not a concern for this upcoming gravitational wave experiment. Note that we ignore the helicity-1 mode because it does not couple to matter, and the quasi-dilaton mode because it does not couple to the tensor mode. Therefore, we can simplify the relevant Lagrangian to \cite{deRham:2014zqa}

\begin{eqnarray}
\label{eq:lag_lin}
   \mathcal{L}&=&\frac{1}{2} h^{\mu\nu}\hat{\mathcal{E}}^{\alpha\beta}_{\mu\nu} h_{\alpha\beta}+\frac{1}{12} \pi \Box \pi - \frac{1}{2}m^2(h^2_{\mu\nu}-h^2)\nonumber\\
   && + \frac{1}{12}m^2\pi h + \frac{1}{6}m^2\pi^2.  
\end{eqnarray}
Corrections to the above expression might come from higher order interactions in the decoupling limit. Generic interactions are of the form \cite{deRham:2014zqa}

\begin{equation}
    \mathcal{L}_{j,k,l}=m^2M_{\rm Pl}^2\Bigg(\frac{h}{M_{\rm Pl}}\Bigg)^j\Bigg(\frac{\partial A}{m M_{\rm Pl}}\Bigg)^{2k}\Bigg(\frac{\partial^2\pi}{m^2M_{\rm Pl}}\Bigg)^l.
\end{equation}
We set $k=0$ for the remainder of the calculation, since we are not interested in helicity-1 interactions. Constraints one should keep in mind are $j+2k+l>2$ and $j,k,l\in N$ \cite{deRham:2014zqa}. If $k=0$, then $j+l>2$. The first interaction that arises at $\Lambda_3$ decoupling scale, $h (\partial^2\pi)^2$, can be removed by field diagonalisation. Therefore, the first correction term is of the form $h(\partial^2\pi)^3$ or $(\partial^2\pi)^4$. We are exploring the possibility of depletion of the tensorial gravitational waves by their decay into scalars, so we examine the first of the two terms. This additional interaction appears in the Lagrangian as
\begin{eqnarray}
    \mathcal{L}_{103}&=&m^2M_{\rm Pl}^2 \Bigg(\frac{h}{M_{\rm Pl}}\Bigg) \Bigg(\frac{\partial^2\pi}{m^2M_{\rm Pl}}\Bigg)^3 \nonumber\\
    &=& \frac{1}{m^4M_{\rm Pl}^2} h(\partial^2\pi)^3.
\end{eqnarray}
Despite a 3-body decay of this type being dynamically forbidden, it is possible in the presence of a background. Since $\Lambda_3^3=m^2M_{\text{Pl}}$, this term reduces to $\mathcal{L}_{103}=\frac{1}{\Lambda_3^6}h(\partial^2\pi)^3$.
The vertex factor contribution to the amplitude is 
\begin{equation}
g_{103}^{\mu\nu} =  \eta^{\mu\nu} \frac{1}{\Lambda_3^6} p_1^2 p_2^2 p_3^2 = \frac{1}{\Lambda_3^6} \eta^{\mu\nu} (m_{\pi}^2)^3,
\end{equation}
where $p_i$s are the outgoing momenta of the scalar particles, and $m_{\pi}$ is the is the mass of the scalar degree of freedom. From (\ref{eq:lag_lin}), we see that $m_{\pi}^2 = \frac{1}{3} m^2$. The resulting amplitude squared is
\begin{equation}
    \sum_{\rm spins} \overline{|M|^2} = 2 \Bigg(\frac{1}{\Lambda_3^6} m_{\pi}^6\Bigg)^2= \frac{2 m_{\pi}^{12}}{m^8 M_{\rm Pl}^4}.
\end{equation}
The expression for the differential decay probability reads
\begin{equation}
   \text{d}\Gamma= \frac{(2\pi)^4}{2m}\sum_{\rm spins} \overline{|M|^2} d\Phi_3(p;p_1,p_2,p_3),
\end{equation}
where $d\Phi_3(p;p_1,p_2,p_3)$ is the phase space of a $1\rightarrow3$ body decay. We apply the treatment of a 3-body decay in the centre of mass frame of the decaying particle found in the Particle Data Group Review \cite{PhysRevD.98.030001}: 
\begin{equation}
    \text{d}\Gamma= \frac{1}{(2\pi)^3}\frac{1}{32m^3}\sum_{\rm spins} \overline{|M|^2} dm_{12}^2 dm_{23}^2,
\end{equation}
where $m$ is the mass of the decaying particle and $m_{ij}^2=p_{ij}^2=(p_i+p_j)^2$ are combinations of masses and momenta of the new particles. Taking the mass of the scalar to be of the order of the mass of the tensor [reasonable assumption from  (\ref{eq:lag_lin})], we find $\int dm_{12}^2 dm_{23}^2 \approx \mathcal{O}(m^4)$. Therefore
\begin{equation}
    \Gamma\approx \frac{1}{(2\pi)^3}\frac{m^5}{M_{Pl}^4}\approx 10^{-225} \text{eV}.
\end{equation}
An order of magnitude estimate implies that a detailed calculation is not worth pursuing. The decay width is too small to affect the signal. We conclude that there can be no observable depletion of the gravitational wave signal due to the decay of the tensor mode into the scalar one.

\subsection{Rotation Curves}

In this section we move from gravitational-wave signals to a regime that allows for astrophysical tests of quasi-dilaton massive gravity \cite{Gannouji:2013rwa}. We are now interested in manifestations of the theory on galactic scales. Quasi-dilaton massive gravity exhibits Vainshtein screening, and within the Vainshtein radius the scalar graviton degree of freedom is heavily suppressed \cite{DAmico:2012hia}. It has been suggested, however, that in beyond Horndeski theories, the Vainshtein screening is only partially effective when time-dependent cosmological fields are considered \cite{Koyama:2015oma}. One would expect a similar phenomenon arising in QDMG. A promising direction of research is analysis of the shape of the galaxy rotation curves. In the following, we first find the evolution of the Bardeen potentials, and then  compare the predictions of the theory to actual data taken by SPARC \cite{Lelli:2016zqa}. Comparing the QDMG predictions to astrophysical data, we set an upper limit to the graviton mass. As mentioned earlier, we work in the decoupling limit of QDMG and use  (\ref{eq:lag_qdmg}) as our starting point.\\

The FLRW metric in the longitudinal gauge reads
\begin{equation}
    ds^2 = a^2(\tau) [-(1+2\Psi(r,\tau)) d\tau^2+(1-2\Phi(r,\tau))\delta_{ij}dx^idx^j],
\end{equation}
with $\Psi$ and $\Phi$ scalar perturbations defined as the usual Bardeen potentials. 
We are concerned with equations of motion of the two Bardeen potentials, as well as of the two galileons. We split  $\pi$ and $\sigma$ respectively into a background cosmological value and its perturbation: $\pi(r,t)=\pi_0(t)+\phi(r,t)$ and $\sigma(r,t)=\sigma_0(t)+\lambda(r,t)$. For the rest of this work, we ignore $a(t)$ and $H(t)$, since we focus on the effects coming solely from $\pi$ and $\sigma$, not from the FLRW metric. Because of this, it is also appropriate to use the decoupling limit of QDMG from (\ref{eq:lag_qdmg}). Furthermore, we ignore time derivatives of all fields and consider only terms up to the cubic galileon \cite{PhysRevD.97.104038}. Higher-order terms containing the fields or their first derivatives are also neglected. Varying the action with respect to the perturbations we then obtain 4 equations in the presence of a non-relativistic source, $T_{\mu\nu}=\text{diag}(\rho,0,0,0)$:
\begin{widetext}
\begin{equation}
\label{eq:eom_phi}
    2\nabla^2\Phi-\frac{\alpha_3+4\alpha_4}{\Lambda^6_3}[(\nabla^2\phi)^3-3\nabla^2\phi(\nabla_i\nabla_j\phi)(\nabla^i\nabla^j\phi)+2(\nabla^i\nabla_j\phi)(\nabla^j\nabla_k\phi)(\nabla^k\nabla_i\phi)] =\frac{1}{M_{\rm Pl}}\rho.
\end{equation}
%\end{widetext}

%\begin{widetext}
\begin{equation}
\label{eq:eom_psi}
    \nabla^2(\Psi-\Phi)=0.
\end{equation}
%\end{widetext}

\begin{equation}
\begin{gathered}
     \omega\nabla^2\lambda - 6 \nabla^2\phi + \frac{2-3\alpha_3}{\Lambda^3_3}[(\nabla^2\phi)^2-(\nabla_i\nabla_j\phi)(\nabla^i\nabla^j\phi)]\\ + \frac{\alpha_3-4\alpha_4}{2\Lambda^6_3} [(\nabla^2\phi)^3-3\nabla^2\phi(\nabla_i\nabla_j\phi)(\nabla^i\nabla^j\phi)+2(\nabla_i\nabla^j\phi)(\nabla_j\nabla^k\phi)(\nabla_k\nabla^i\phi)] =0
\end{gathered}
\end{equation}

\begin{equation}
\begin{gathered}
     \frac{3}{2}\nabla^2\phi +\frac{3}{4}\frac{(2+3\alpha_3)}{\Lambda^3_3}[(\nabla^2\phi)^2-(\nabla_i\nabla_j\phi)(\nabla^i\nabla^j\phi)]- \frac{3(\alpha_3+4\alpha_4)}{\Lambda^6_3}[(\nabla^2\Psi)(\nabla^2\phi)^2-(\nabla^2\Psi)( \nabla_i\nabla_j\phi)(\nabla^i\nabla^j\phi) \\ - 2 (\nabla^2\phi)( \nabla_i\nabla_j\phi)(\nabla^i\nabla^j\Psi)+2(\nabla_i\nabla^j\Psi)(\nabla_j\nabla^k\phi)(\nabla_k\nabla^i\phi)]-6\nabla^2\lambda+\frac{2(2-3\alpha_3)}{\Lambda^3_3}[(\nabla^2\lambda)(\nabla^2\phi)\\-( \nabla_i\nabla_j\lambda)(\nabla^i\nabla^j\phi)]+\frac{3(\alpha_3-4\alpha_4)}{2\Lambda^6_3}[(\nabla^2\lambda)(\nabla^2\phi)^2-(\nabla^2\lambda)(\nabla_i\nabla_j\phi)(\nabla^i\nabla^j\phi)-2(\nabla^2\phi)( \nabla_i\nabla_j\phi)(\nabla^i\nabla^j\lambda) \\+2(\nabla_i\nabla^j\lambda)(\nabla_j\nabla^k\phi)(\nabla_k\nabla^i\phi)]=\frac{\rho}{2M_{\rm Pl}}.
\end{gathered}
\end{equation}
\end{widetext}

It is straight-forward to notice that the typical Poisson's equation for the Bardeen potential $\Phi$ is altered by the presence of the scalar mode shown in  (\ref{eq:eom_phi}). Interestingly, however, the Laplace's equation (\ref{eq:eom_psi}) remains the same as in the GR case. Already at the level of the equations of motion, we can deduce that gravitational lensing tests are not an appropriate means of constraining this theory, since the quantity $\frac{\Phi+\Psi}{2\Phi}=1$ is indistinguishable from the exact same GR prediction.\\

Assuming spherical symmetry, integrating the above expressions by parts and using variables
\begin{equation}
     x \equiv \frac{\phi'}{r},\; y \equiv \frac{\Psi'}{r},\; z \equiv \frac{\Phi'}{r},\; A \equiv \frac{M(r)}{8\pi M_{\rm Pl}r^3},\; q \equiv \frac{\lambda'}{r},
\end{equation}
where the primes denote radial derivatives, we obtain the following set of simultaneous equations:

\begin{equation}
\label{zxA}
     z - \frac{\alpha_3+4\alpha_4}{\Lambda^6_3} x^3= A,
\end{equation}

\begin{equation}
\label{yz}
     y - z = 0,
\end{equation}

\begin{equation}
     \omega q - 6x + \frac{2(2-3\alpha_3)}{\Lambda^3_3}x^2 + \frac{\alpha_3-4\alpha_4}{\Lambda^6_3}x^3=0,
\end{equation}

\begin{equation}
\begin{gathered}
\label{eq:eom_x}
     \frac{3}{2}x + \frac{3(2+3\alpha_3)}{2\Lambda^3_3}x^2 - \frac{6(\alpha_3+4\alpha_4)}{\Lambda^6_3}x^2y\\- 6q + \frac{4(2-3\alpha_3)}{\Lambda^3_3} qx + \frac{3(\alpha_3-4\alpha_4)}{\Lambda^6_3} x^2 q = A.
     \end{gathered}
\end{equation}
Combining the first 3 coupled equations and plugging into (\ref{eq:eom_x}) gives a quintic for $x$, for which we give the full expression in the Appendix. Taking the quintic term to be the dominant one, we can approximate 

\begin{equation}
\label{eqx}
    - \frac{3}{\Lambda^{12}_3}\Bigg[ 2(\alpha_3+4\alpha_4)^2+\frac{1}{\omega}(\alpha_3-4\alpha_4)^2\Bigg] x^5 = A.
\end{equation}
It is this equation that we will use to find the form of $x$, which will then provide a prediction for the shape of rotation curves.

The velocity of objects within the galaxy undergoing circular motion is 

\begin{equation}
     \frac{v^2}{r} = \frac{\text{d}\Psi}{\text{d}r},
\end{equation}
which in terms of our new variables reads

\begin{equation}
     v^2 = r^2  y.
\end{equation}
Most of galaxy's mass is in its dark matter halo, and in this work we assume it to obey the Navarro-Frenk-White (NFW) profile \cite{Navarro:1995iw}:
\begin{equation}
    \rho_{\rm NFW} = \frac{\rho_{\rm s}}{\frac{r}{r_{\rm s}} \Big(1+ \frac{r}{r_{\rm s}}\Big)^2},
\end{equation}{}
with $r_{\rm s}$ and $\rho_s$ as the typical halo parameters.\\

Using \eqref{zxA}, \eqref{yz} and \eqref{eqx}, and assuming a Navarro-Frenk-White distribution, we get the following equation for the dark matter velocity profile 
\begin{equation}
\begin{gathered}
\label{eqvqdmg}
   v_{\rm dm}^2= 4\pi G r_{\rm s}^2 \rho_{\rm s} \Bigg[\frac{1}{R}\Big[\text{ln}\Big(1+R\Big)-\Big(1+\frac{1}{R}\Big)^{-1}\Big]\\ -\boldsymbol{\gamma} \: R^{\frac{1}{5}} \Big[\text{ln}\Big(1+R\Big)-\Big(1+\frac{1}{R}\Big)^{-1}\Big]^{\frac{3}{5}}\Bigg],
\end{gathered}
\end{equation}
where $R = \frac{r}{r_{\rm s}}$, with 
\begin{equation}
  \boldsymbol{\gamma}=\big(4\pi G \rho_{\rm s}\big)^{-\frac{2}{5}} m^{\frac{4}{5}} \frac{\alpha_3+4\alpha_4}{[6(\alpha_3+4\alpha_4)^2+\frac{3}{\omega}(\alpha_3-4\alpha_4)^2]^{\frac{3}{5}}}.
\end{equation}{}
It is this parameter, $\boldsymbol{\gamma}$, that we can constrain by fitting the theoretical predictions to the rotation curves data. We use best-fit values of $\omega$ and $\alpha$ coefficients from \cite{Gannouji:2013rwa}. Since $\boldsymbol{\gamma}$ is a function of $m$, by constraining $\boldsymbol{\gamma}$ we can put an upper limit on the graviton mass.\\

We use the data from the SPARC galaxy catalogue \cite{Lelli:2016zqa} to reconstruct rotation curves. The observed speed is not only due to the dark matter, but there are gas, disk and bulge (if applicable) contributions too:

\begin{eqnarray}
    v^2(r)&=& v_{\rm gas}^2(r)  + \Upsilon_{\rm disk}\, v_{\rm disk}^2(r) \nonumber\\
    &&+ \Upsilon_{\rm bulge}\,  v_{\rm bulge}^2(r)  + v_{\rm dm}^2(r),
\end{eqnarray}
where $\Upsilon$ is the stellar-to-mass ratio. In other words, we must subtract all the matter contributions from the data to obtain $v_{\rm dm}$, and then compare against our predictions. We use $\Upsilon_{\rm disk}, \Upsilon_{\rm bulge}, r_{\rm s}$ and $\rho_{\rm s}$ values from Table 4 in \cite{deAlmeida:2018kwq}. We are aware that a full treatment would require a Monte Carlo Markov Chain simulation \cite{10.1093/mnras/stw3101} to find posterior best-fit values of all the parameters: in this case the NFW parameters and the $\boldsymbol{\gamma}$ parameter from QDMG, as well as $\Upsilon_{\rm disk}$ and $\Upsilon_{\rm bulge}$. We leave this approach for future work. \\

All of the galaxies in the SPARC catalogue are characterised by a quantity, Q, which refers to the quality of the galaxy's rotation curve. We consider galaxies with Q=1, in other words galaxies with the best quality rotation curves. We also choose carefully high mass and high luminosity galaxies because the NFW profile provides the best fit for those type of galaxies, \cite{10.1093/mnras/stw3101}. On a single plot (see for instance the case of 2 representative galaxies in Figure~1) we compare the NFW profile in the GR case and the corresponding one in the QDMG case fit for the data.
The galaxies that we take show consistency in the value of $\boldsymbol{\gamma}$ that fits the dark matter galaxy profile (\ref{eqvqdmg}) to data. The constraint placed on the graviton mass from this is 

\begin{equation}
\label{eq:bound}
m \leq 10^{-31} \, \text{eV}.
\end{equation}

This bound satisfies all constraints hitherto imposed on the mass of the graviton from LIGO/VIRGO and Solar System tests \cite{2018CQGra..35qLT01W}. Typically, massive gravity theories are motivated with the aim of explaining the origin of Dark Energy, and indeed values of $m$ of the order of Hubble, $m \approx 10^{-33}$ eV, can accomplish this \cite{deRham:2010kj}. Our constraint does not disqualify such a {\sl statement}, neither does it rule out massive gravity as a proposal for explaining the late-time accelerated expansion of the Universe. However, one may wonder whether one can come up with an astrophysical test that could falsify massive gravity. In other words, could such a tiny mass be ever detected? The most promising constraints we can hope to probe in the upcoming decades will be through LISA when we will be able to detect masses up to $10^{-25}$ eV \cite{2018CQGra..35qLT01W}. At present, we do not see a way for the constraint derived in (\ref{eq:bound}) to be tested with gravitational waves data.\\

However, one should keep in mind that our result comes with subtleties and caveats. Along the way, we made numerous approximations: we decided to take (scientifically motivated) limits of QDMG that allowed us to proceed with our analytical analysis. These approximations included going only up to the cubic galileon in the decoupling limit, as well as setting the background values of the scalar fields $\pi_0$ and $\sigma_0$ to 0.  We anticipate opportunities for future work within going beyond these approximations.

\begin{figure}[h]
    \centering
    \begin{minipage}{0.45\textwidth}
        \centering
        \includegraphics[width=0.9
        \textwidth]{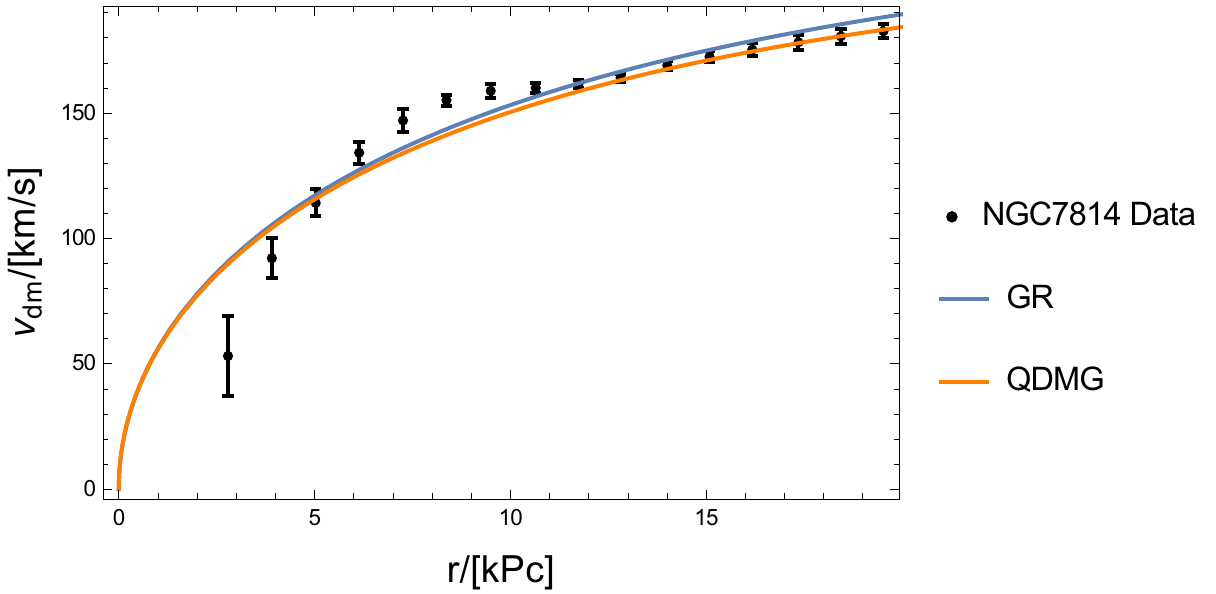} 
      \vspace{1.cm}
    \end{minipage}\hfill
    \begin{minipage}{0.45\textwidth}
        \centering
        \includegraphics[width=0.9\textwidth]{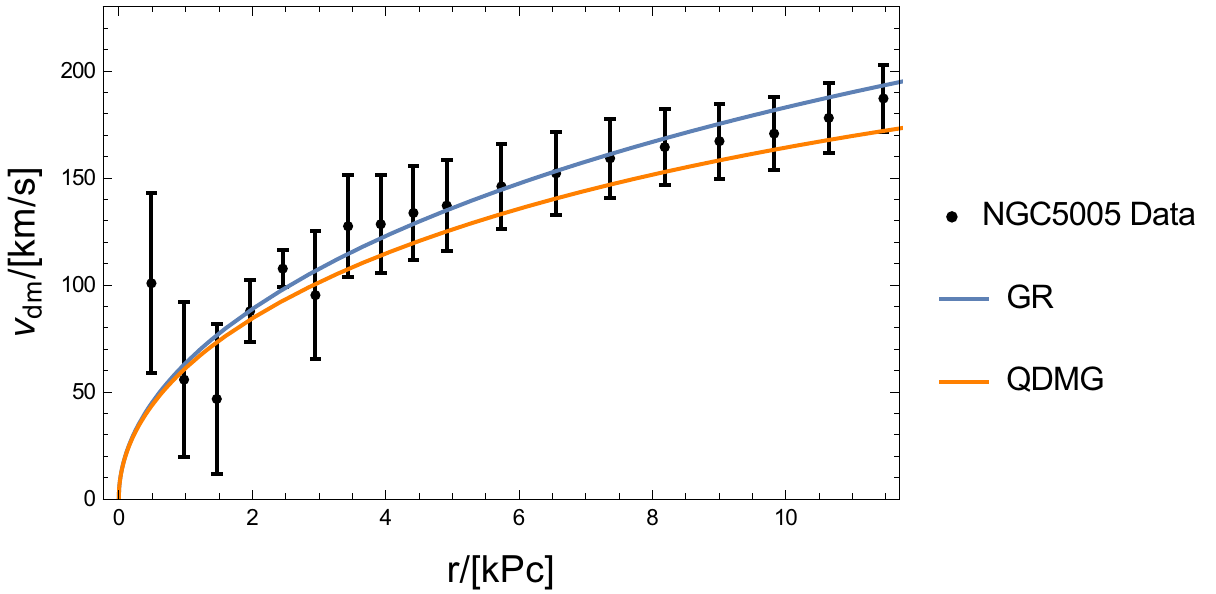} % second figure itself
        \caption{Dark Matter contribution to rotation curves, GR case and QDMG case, for 2 galaxies in the SPARC catalogue, NGC7814 (top panel) and NGC5005 (bottom panel). In the limit $m=0$, the QDMG fit coincides with the GR fit. We have investigated what maximum graviton mass still keeps the theoretical QDMG fit within the error bars of the data.}
   \end{minipage}
    \label{fig:plots}
\end{figure}

\section{Discussion}

We have looked for ways to lift the mathematical success of dRGT theory to a phenomenological level. The lack of stable cosmological solutions in massive gravity forced us to extend the theory by adding a quasi-dilaton field, leading to quasi-dilaton massive gravity. Having resorted to this theory, we found a promising direction of study in analysing rotation curves of galaxies. The new scalar degrees of freedom in QDMG affect equations of motion of the dark matter within the galaxies and they are expected to alter the dark matter halo profile. The extent to which the dark matter profile can change is constrained by rotation curves data. We combine the observations collected in the SPARC database with our theoretical prediction. In order for QDMG to agree with the data, the graviton mass must be  $m \leq 10^{-31} \, \text{eV}$. This result does not contradict any previous bounds on the mass of the graviton, and leaves the massive gravity as a viable dark energy candidate theory. The next step would be to perform a MCMC simulation and obtain a best-fit value of graviton mass, along with the NFW parameters, rather than take the NFW parameters from a MCMC simulation in the GR case, and then constrain $m$. This would lead to a more exhaustive bound on the mass. Additionally, one could expand to include the background fields and go up to the quartic galileon.

%\newpage
\begin{acknowledgements}
The authors would like to thank Luca Amendola and Claudia de Rham for useful discussions.
K.M. is supported by King’s College London through a Postgraduate International 
Scholarship.
M.S. is supported in part by the Science and Technology Facility Council (STFC), United Kingdom,
under the research grant ST/P000258/1.
\end{acknowledgements}

\section*{Appendix}

\subsection*{QDMG Lagrangian in the decoupling limit}

We write below all terms in the Lagrangian of quasi-dilaton massive gravity in the decoupling limit,
\begin{widetext}
\begin{equation}
\begin{gathered}
    \mathcal{L}^{\text{DL}}_{\text{QDMG}} = - \frac{1}{4}\Big( h^{\mu\nu}\hat{\mathcal{E}}^{\alpha\beta}_{\mu\nu}h_{\alpha\beta}+\sum^5_{n=2} \frac{c_n}{\Lambda^{3(n-2)}_3} \mathcal{L}^{(n)}_{\text{Gal}}[\pi] - \frac{2(\alpha_3+4\alpha_4)}{\Lambda^6_3}h^{\mu\nu}X^{(3)}_{\mu\nu}[\Pi]\Big)  
    - \frac{\omega}{2} (\partial \sigma)^2 \\+ \frac{1}{2} \sigma \sum^4_{n=1}\frac{(4-n)\alpha_n-(n+1)\alpha_{n+1}}{\Lambda^{3(n-1)}_3} \mathcal{L}_n[\Pi]   + \frac{1}{2 M_{\text{Pl}}} h_{\mu\nu}T^{\mu\nu}+ \frac{1}{2 M_{\text{Pl}}}  \pi T- \frac{2+3\alpha_3}{4M_{\text{Pl}} \Lambda^3_3} \partial_{\mu}\pi \partial_{\nu}\pi T^{\mu\nu}.
\end{gathered}
\end{equation}
\end{widetext}

The galileon Lagrangians are defined below:

\begin{equation}
    \mathcal{L}^{(2)}_{\text{Gal}}[\pi] = (\partial \pi)^2,
\end{equation}{}

\begin{equation}
   \mathcal{L}^{(3)}_{\text{Gal}}[\pi] = (\partial \pi)^2 \Box \pi,
\end{equation}{}

\begin{equation}
  \mathcal{L}^{(4)}_{\text{Gal}}[\pi]  = (\partial \pi)^2 \Big((\Box \pi)^2 - \nabla_{\mu}\nabla_{\nu}\pi\nabla^{\mu}\nabla^{\nu}\pi\Big),
\end{equation}{}{}

\begin{equation}
\begin{gathered}
    \mathcal{L}^{(5)}_{\text{Gal}}[\pi] = (\partial \pi)^2 \Big( (\Box \pi)^3 - 3 \Box \pi \nabla_{\mu}\nabla_{\nu}\pi\nabla^{\mu}\nabla^{\nu}\pi\\ + 2 \nabla^{\mu}\nabla_{\nu}\pi\nabla^{\nu}\nabla_{\lambda}\pi \nabla^{\lambda}\nabla_{\mu}\pi    \Big).
\end{gathered}
\end{equation}{}

Next, we define $X^{(3)}_{\mu\nu}$ as

\begin{equation}
\begin{gathered}
   X^{(3)}_{\mu\nu} = \Big([\Pi]^3 - 3[\Pi][\Pi^2] + 2[\Pi^3]\Big) \eta_{\mu\nu}\\ -3\Big( [\Pi]^2 \Pi_{\mu\nu} - 2 [\Pi] \Pi^2_{\mu\nu} - [\Pi^2] \Pi_{\mu\nu} + 2 \Pi^3_{\mu\nu}  \Big) ,
\end{gathered}
\end{equation}{}
where $\Pi_{\mu\nu}=\partial_{\mu}\partial_{\nu}\pi$.\\

Finally, we explicitly write out the expressions for $\mathcal{L}_n[\Pi]$: 

\begin{equation}
   \mathcal{L}_1[\Pi] = 3![\Pi],
\end{equation}{}

\begin{equation}
   \mathcal{L}_2[\Pi] = 2 \Big([\Pi]^2 - [\Pi^2]  \Big),
\end{equation}{}

\begin{equation}
   \mathcal{L}_3[\Pi] = [\Pi]^3 - 3[\Pi][\Pi^2] + 2[\Pi^3].
\end{equation}{}

\vskip1truecm

\subsection*{Equation for $x$}

We expand and show in full detail the equation for $x=\frac{\phi'}{r}$:
\begin{widetext}
\begin{equation}
\begin{gathered}
   \Bigg[\frac{3}{2}-\frac{36}{\omega}\Bigg]x + \frac{3}{\Lambda^3_3}\Bigg[\frac{2+3\alpha_3}{2}+\frac{12(2-3\alpha_3)}{\omega}-\frac{2(\alpha_3+4\alpha_4)}{\Lambda^3_3} A\Bigg]x^2 + \frac{8}{\Lambda^6_3}\Bigg[\frac{3(\alpha_3-4\alpha_4)-(2-3\alpha_3)^2}\omega\Bigg]x^3 \\-\frac{10}{\Lambda^9_3}\Bigg[\frac{(2-3\alpha_3)(\alpha_3-4\alpha_4)}{\omega}\Bigg]x^4 -\frac{3}{\Lambda^{12}_3}\Bigg[ 2(\alpha_3+4\alpha_4)^2+\frac{1}{\omega}(\alpha_3-4\alpha_4)^2\Bigg] x^5 = A    
\end{gathered}
\end{equation}
 \end{widetext}
 
\newpage
 \let\cleardoublepage\clearpage
 \bibliographystyle{unsrt} %bibliography style
 \bibliography{qdmg} %References file

\begin{thebibliography}{10}

\bibitem{Reynaud:2008yd}
Serge Reynaud and Marc-Thierry Jaekel.
\newblock {Tests of general relativity in the solar system}.
\newblock {\em Proc. Int. Sch. Phys. Fermi}, 168:203--217, 2009.

\bibitem{Collett:2018gpf}
Thomas~E. Collett, Lindsay~J. Oldham, Russell~J. Smith, Matthew~W. Auger,
  Kyle~B. Westfall, David Bacon, Robert~C. Nichol, Karen~L. Masters, Kazuya
  Koyama, and Remco van~den Bosch.
\newblock {A precise extragalactic test of General Relativity}.
\newblock {\em Science}, 360:1342, 2018.

\bibitem{TheLIGOScientific:2016src}
B.~P. Abbott et~al.
\newblock {Tests of general relativity with GW150914}.
\newblock {\em Phys. Rev. Lett.}, 116(22):221101, 2016.
\newblock [Erratum: Phys. Rev. Lett.121,no.12,129902(2018)].

\bibitem{Kramer:2006nb}
M.~Kramer et~al.
\newblock {Tests of general relativity from timing the double pulsar}.
\newblock {\em Science}, 314:97--102, 2006.

\bibitem{PhysRevD.71.063513}
Sean~M. Carroll, Antonio De~Felice, Vikram Duvvuri, Damien~A. Easson, Mark
  Trodden, and Michael~S. Turner.
\newblock Cosmology of generalized modified gravity models.
\newblock {\em Phys. Rev. D}, 71:063513, Mar 2005.

\bibitem{TheLIGOScientific:2017qsa}
B.~P. Abbott et~al.
\newblock {GW170817: Observation of Gravitational Waves from a Binary Neutron
  Star Inspiral}.
\newblock {\em Phys. Rev. Lett.}, 119(16):161101, 2017.

\bibitem{Monitor:2017mdv}
B.~P. Abbott et~al.
\newblock {Gravitational Waves and Gamma-rays from a Binary Neutron Star
  Merger: GW170817 and GRB 170817A}.
\newblock {\em Astrophys. J.}, 848(2):L13, 2017.

\bibitem{Baker:2017hug}
T.~Baker, E.~Bellini, P.~G. Ferreira, M.~Lagos, J.~Noller, and I.~Sawicki.
\newblock {Strong constraints on cosmological gravity from GW170817 and GRB
  170817A}.
\newblock {\em Phys. Rev. Lett.}, 119(25):251301, 2017.

\bibitem{Ezquiaga:2017ekz}
Jose~María Ezquiaga and Miguel Zumalacárregui.
\newblock {Dark Energy After GW170817: Dead Ends and the Road Ahead}.
\newblock {\em Phys. Rev. Lett.}, 119(25):251304, 2017.

\bibitem{PhysRevLett.119.251303}
Jeremy Sakstein and Bhuvnesh Jain.
\newblock Implications of the neutron star merger gw170817 for cosmological
  scalar-tensor theories.
\newblock {\em Phys. Rev. Lett.}, 119:251303, Dec 2017.

\bibitem{deRham:2010kj}
Claudia de~Rham, Gregory Gabadadze, and Andrew~J. Tolley.
\newblock {Resummation of Massive Gravity}.
\newblock {\em Phys. Rev. Lett.}, 106:231101, 2011.

\bibitem{DAmico:2012hia}
Guido D'Amico, Gregory Gabadadze, Lam Hui, and David Pirtskhalava.
\newblock {Quasidilaton: Theory and cosmology}.
\newblock {\em Phys. Rev.}, D87:064037, 2013.

\bibitem{vanDam:1970vg}
H.~van Dam and M.~J.~G. Veltman.
\newblock {Massive and massless Yang-Mills and gravitational fields}.
\newblock {\em Nucl. Phys.}, B22:397--411, 1970.

\bibitem{Zakharov:1970cc}
V.~I. Zakharov.
\newblock {Linearized gravitation theory and the graviton mass}.
\newblock {\em JETP Lett.}, 12:312, 1970.
\newblock [Pisma Zh. Eksp. Teor. Fiz.12,447(1970)].

\bibitem{Babichev:2013usa}
Eugeny Babichev and Cédric Deffayet.
\newblock {An introduction to the Vainshtein mechanism}.
\newblock {\em Class. Quant. Grav.}, 30:184001, 2013.

\bibitem{Hinterbichler:2011tt}
Kurt Hinterbichler.
\newblock {Theoretical Aspects of Massive Gravity}.
\newblock {\em Rev. Mod. Phys.}, 84:671--710, 2012.

\bibitem{PhysRevD.83.103516}
Claudia de~Rham, Gregory Gabadadze, Lavinia Heisenberg, and David Pirtskhalava.
\newblock Cosmic acceleration and the helicity-0 graviton.
\newblock {\em Phys. Rev. D}, 83:103516, May 2011.

\bibitem{Cheung:2016yqr}
Clifford Cheung and Grant~N. Remmen.
\newblock {Positive Signs in Massive Gravity}.
\newblock {\em JHEP}, 04:002, 2016.

\bibitem{deRham:2014zqa}
Claudia de~Rham.
\newblock {Massive Gravity}.
\newblock {\em Living Rev. Rel.}, 17:7, 2014.

\bibitem{PhysRevD.99.104032}
Michael Kenna-Allison, A.~Emir G\"umr\"uk\ifmmode \mbox{\c{c}}\else
  \c{c}\fi{}\"uo\ifmmode~\check{g}\else \v{g}\fi{}lu, and Kazuya Koyama.
\newblock Viability of bigravity cosmology.
\newblock {\em Phys. Rev. D}, 99:104032, May 2019.

\bibitem{Cai:2014upa}
Yi-Fu Cai and Emmanuel~N. Saridakis.
\newblock {Cosmology of F(R) nonlinear massive gravity}.
\newblock {\em Phys. Rev.}, D90(6):063528, 2014.

\bibitem{Damour:1994zq}
T.~Damour and Alexander~M. Polyakov.
\newblock {The String dilaton and a least coupling principle}.
\newblock {\em Nucl. Phys.}, B423:532--558, 1994.

\bibitem{Ondo:2013wka}
Nicholas~A. Ondo and Andrew~J. Tolley.
\newblock {Complete Decoupling Limit of Ghost-free Massive Gravity}.
\newblock {\em JHEP}, 11:059, 2013.

\bibitem{Dar:2018dra}
Furqan Dar, Claudia De~Rham, J.~Tate Deskins, John~T. Giblin, and Andrew~J.
  Tolley.
\newblock {Scalar Gravitational Radiation from Binaries: Vainshtein Mechanism
  in Time-dependent Systems}.
\newblock {\em Class. Quant. Grav.}, 36(2):025008, 2019.

\bibitem{Gannouji:2013rwa}
Radouane Gannouji, Md.~Wali Hossain, M.~Sami, and Emmanuel~N. Saridakis.
\newblock {Quasidilaton nonlinear massive gravity: Investigations of background
  cosmological dynamics}.
\newblock {\em Phys. Rev.}, D87:123536, 2013.

\bibitem{Jain:2015edg}
Rajeev~Kumar Jain, Chris Kouvaris, and Niklas~Grønlund Nielsen.
\newblock {White Dwarf Critical Tests for Modified Gravity}.
\newblock {\em Phys. Rev. Lett.}, 116(15):151103, 2016.

\bibitem{Koyama:2015oma}
Kazuya Koyama and Jeremy Sakstein.
\newblock {Astrophysical Probes of the Vainshtein Mechanism: Stars and
  Galaxies}.
\newblock {\em Phys. Rev.}, D91:124066, 2015.

\bibitem{Salzano:2017qac}
Vincenzo Salzano, David~F. Mota, Salvatore Capozziello, and Megan Donahue.
\newblock {Breaking the Vainshtein screening in clusters of galaxies}.
\newblock {\em Phys. Rev.}, D95(4):044038, 2017.

\bibitem{Gumrukcuoglu:2011zh}
A.~Emir Gumrukcuoglu, Chunshan Lin, and Shinji Mukohyama.
\newblock {Cosmological perturbations of self-accelerating universe in
  nonlinear massive gravity}.
\newblock {\em JCAP}, 1203:006, 2012.

\bibitem{LIGOScientific:2019fpa}
B.~P. Abbott et~al.
\newblock {Tests of General Relativity with the Binary Black Hole Signals from
  the LIGO-Virgo Catalog GWTC-1}.
\newblock 2019.

\bibitem{deRham:2016nuf}
Claudia de~Rham, J.~Tate Deskins, Andrew~J. Tolley, and Shuang-Yong Zhou.
\newblock {Graviton Mass Bounds}.
\newblock {\em Rev. Mod. Phys.}, 89(2):025004, 2017.

\bibitem{Creminelli:2018xsv}
Paolo Creminelli, Matthew Lewandowski, Giovanni Tambalo, and Filippo Vernizzi.
\newblock {Gravitational Wave Decay into Dark Energy}.
\newblock {\em JCAP}, 1812(12):025, 2018.

\bibitem{deRham:2012fw}
Claudia de~Rham, Andrew~J. Tolley, and Daniel~H. Wesley.
\newblock {Vainshtein Mechanism in Binary Pulsars}.
\newblock {\em Phys. Rev.}, D87(4):044025, 2013.

\bibitem{deRham:2018red}
Claudia de~Rham and Scott Melville.
\newblock {Gravitational Rainbows: LIGO and Dark Energy at its Cutoff}.
\newblock {\em Phys. Rev. Lett.}, 121(22):221101, 2018.

\bibitem{PhysRevD.98.030001}
Particle~Data Group.
\newblock Review of particle physics.
\newblock {\em Phys. Rev. D}, 98:030001, Aug 2018.

\bibitem{Lelli:2016zqa}
Federico Lelli, Stacy~S. McGaugh, and James~M. Schombert.
\newblock {SPARC: Mass Models for 175 Disk Galaxies with Spitzer Photometry and
  Accurate Rotation Curves}.
\newblock {\em Astron. J.}, 152:157, 2016.

\bibitem{PhysRevD.97.104038}
Shun Arai and Atsushi Nishizawa.
\newblock Generalized framework for testing gravity with gravitational-wave
  propagation. ii. constraints on horndeski theory.
\newblock {\em Phys. Rev. D}, 97:104038, May 2018.

\bibitem{Navarro:1995iw}
Julio~F. Navarro, Carlos~S. Frenk, and Simon D.~M. White.
\newblock {The Structure of cold dark matter halos}.
\newblock {\em Astrophys. J.}, 462:563--575, 1996.

\bibitem{deAlmeida:2018kwq}
Álefe O.~F. de~Almeida, Luca Amendola, and Viviana Niro.
\newblock {Galaxy rotation curves in modified gravity models}.
\newblock {\em JCAP}, 1808(08):012, 2018.

\bibitem{10.1093/mnras/stw3101}
Harley Katz, Federico Lelli, Stacy~S. McGaugh, Arianna Di~Cintio, Chris~B.
  Brook, and James~M. Schombert.
\newblock {Testing feedback-modified dark matter haloes with galaxy rotation
  curves: estimation of halo parameters and consistency with $\Lambda$CDM
  scaling relations}.
\newblock {\em Monthly Notices of the Royal Astronomical Society},
  466(2):1648--1668, 12 2016.

\bibitem{2018CQGra..35qLT01W}
Clifford~M. {Will}.
\newblock {Solar system versus gravitational-wave bounds on the graviton mass}.
\newblock {\em Classical and Quantum Gravity}, 35(17):17LT01, Sep 2018.

\end{thebibliography}

\end{document}